# Evidence for an effective gauge field in flakes of superconducting 4Hb-$TaS_2$

Shalev Gur[1,2], T.R. Devidas[1,2], Eylon Persky[1,2], Roni Anna Gofman[1,2], Irena Feldman[3], Avior Almoalem[3], Sudip Kumar Saha[1], Mark H. Fischer[4], Avraham Klein[1], Jonathan Ruhman[1], Amit Kanigel[3] & Beena Kalisky[1,2✉]

[1]Department of Physics, Bar Ilan University, Ramat Gan, Israel. [2]Bar Ilan Institute of Nanotechnology and Advanced materials, Bar Ilan University, Ramat Gan, Israel. [3]Department of Physics, Technion–Israel Institute of Technology, Haifa, Israel. [4]Department of Physics, University of Zurich, Zurich, Switzerland
✉e–mail: beena@biu.ac.il

**Understanding the normal state of unconventional superconductors is central to uncovering the origin of superconductivity itself. Unconventional superconductivity often emerges from correlated electronic states characterized by intertwined symmetry-breaking orders. A striking example is 4Hb-$TaS_2$, whose superconducting state exhibits a remarkable range of phenomena, including an unusual magnetic memory effect. Previous experiments showed that magnetic memory can be encoded above $T_c$ and later revealed through spontaneous vortices below $T_c$, suggesting time-reversal symmetry breaking. However, since memory and vorticity are intertwined, this prevents establishing whether the effect results from trapped flux, weak conventional magnetism, or a true hidden electronic order. Here, we show that magnetic memory persists in the vortex-free Meissner regime of exfoliated 4Hb-$TaS_2$ flakes, where it appears as a stable remanent modification of the superconducting screening response. Our results rule out explanations based on trapped vortices. Furthermore, the spatial profile of the flux generated by trained samples at zero field is inconsistent with conventional weak magnetism, suggesting a hidden electronic phase that imprints an effective flux on the superconducting condensate.**

Memory effects in superconductors are commonly associated with vortex physics. In type-II superconductors, magnetic history dependence arises from vortex pinning, flux trapping, and critical-state behavior, producing remanent magnetization and hysteretic responses[1,2]. However, unconventional superconductors may also exhibit memory effects associated with internal electronic degrees of freedom. In particular, time-reversal-symmetry-breaking phases can retain information about magnetic and thermal history through the selection of discrete configurations of the underlying order parameter[3–8]. Such memory may arise from weak magnetic order that remains difficult to detect directly, or from more exotic electronic states that couple to the superconducting condensate through effective gauge fields[9–11].

4Hb-$TaS_2$ provides a unique platform for studying such a phenomenon. 4Hb-$TaS_2$ is a layered van der Waals heterostructure composed of alternating 1T-$TaS_2$ and 2H-$TaS_2$ layers[12]. Bulk 1T-$TaS_2$ layers host a commensurate charge-density-wave state and form a Mott insulating phase on a triangular lattice, which has been proposed as a candidate quantum spin liquid[13,14]. In contrast, 2H-$TaS_2$ is a superconductor with coexisting charge-density-wave order[15]. The 4Hb structure exhibits enhanced

superconductivity with a transition temperature $T_c \approx 2.7$ K, together with signatures of unconventional superconductivity including nematicity, chirality, and multi-component order[16–19].

A recent experiment reported a magnetic memory effect associated with a characteristic temperature $T_* = 3.6$ K[20]. In that study, cooling through $T_*$ in a magnetic training field produced no detectable magnetic signal after turning the field off in the normal state. However, upon subsequently entering the superconducting phase, spontaneous vortices emerged with a polarity determined by the training field. These observations show that the magnetic memory was encoded above $T_c$ and later revealed through superconductivity.

There are several possible explanations for the reported magnetic memory. First, the observed vortices could result from trapped flux, vortex pinning, or rare defects that preferentially nucleate vortices[21–23]. Second, an extremely weak time-reversal-symmetry-breaking magnetic order below the sensitivity of conventional probes in the normal state is amplified upon entering the superconducting state. Finally, the memory could originate from a hidden electronic phase that generates an effective flux for superconducting electrons while producing little or no detectable magnetic field. Distinguishing between these scenarios requires, first, accessing the memory independently of vortex formation and determining whether it is associated with an ordinary magnetic field.

To address this challenge, we investigate magnetic memory in thin exfoliated 4Hb-$TaS_2$ flakes using spatially resolved scanning SQUID magnetometry and susceptometry[24,25]. In contrast to bulk crystals, where residual trapped vortices and spatially inhomogeneous vortex distributions complicate magnetic-memory interpretation[20], thin flakes enable measurements in the Meissner regime. In these samples, the London penetration depth remains small relative to the lateral dimensions of the flake, producing well-defined edge-localized screening currents and a predictable Meissner response while strongly suppressing vortex trapping. This geometry allows the magnetic memory to be probed directly through its influence on reversible Meissner screening currents.

Remarkably, we find a robust memory effect even in the vortex-free flakes, with similar strength to the previous measurements on bulk samples. A flake trained by an external field exhibits a remanent flux response even after warming above $T_c$ and recooling in zero field. The memory displays pronounced hysteresis under both temperature and training-field evolution, demonstrating that it originates from an underlying hidden phase that breaks time-reversal symmetry. Even more strikingly, the spatial profile of the remanent signal is nearly identical to that generated by an applied magnetic field, differing only by a temperature-dependent scale factor. The persistence of memory in the Meissner regime rules out mechanisms based on trapped vortices and vortex nucleation. Detailed analysis of the spatial response excludes ordinary magnetic fields, including weak magnetization amplified by superconductivity, as the source of the memory. Our results therefore provide evidence for a hidden electronic phase that acts on the superconducting condensate through an effective, non-magnetic flux, generating a remanent Meissner response while remaining below the sensitivity of conventional magnetic and thermodynamic probes.

## Existence of a memory effect in the Meissner response

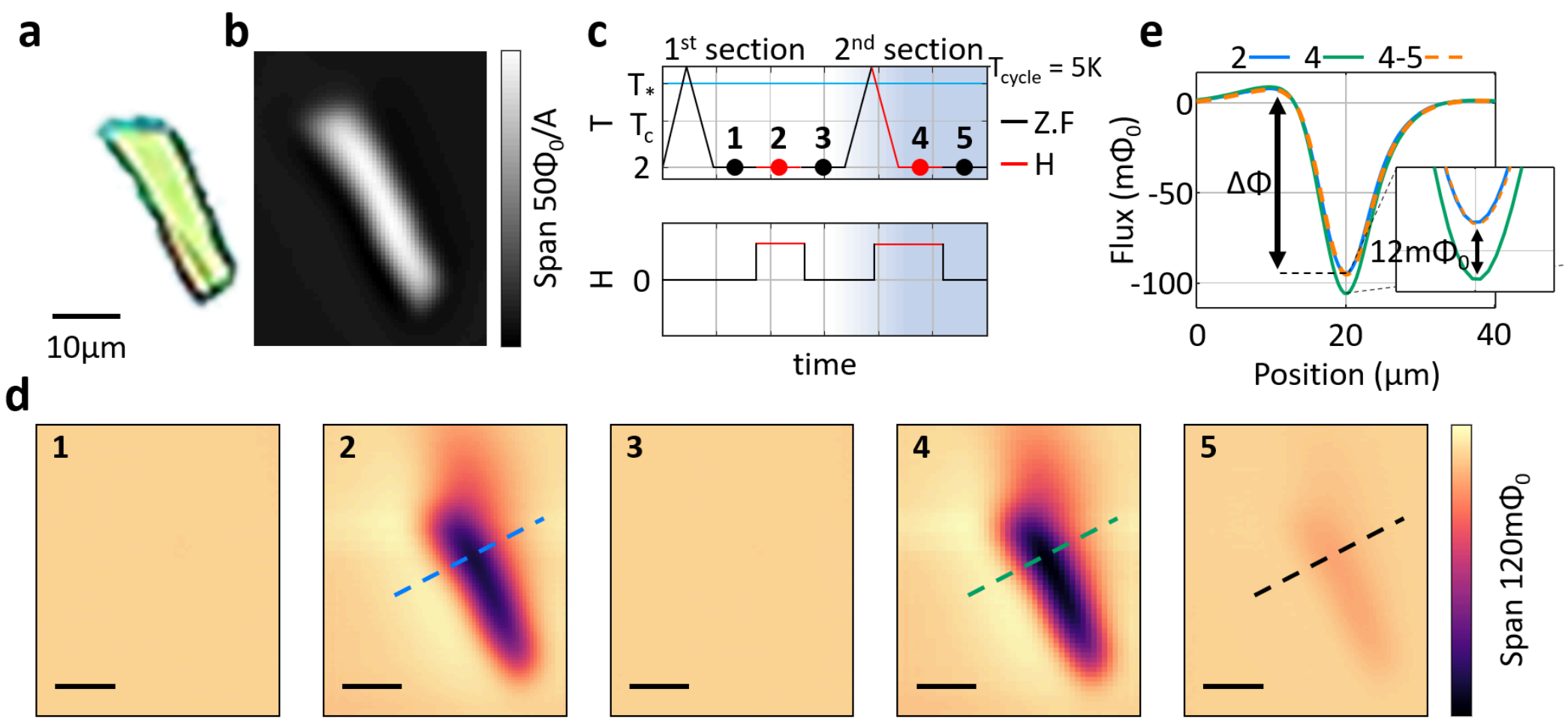

**Fig. 1 | Remanent flux revealed by a ZFC/FC protocol. a** Optical micrograph of a representative 4Hb-$TaS_2$ flake measured in this study. **b** Scanning SQUID susceptometry (χ) image of the same flake, used to determine the flake boundaries and sensor position. The grayscale bar indicates the susceptometry signal range. **c** Schematic of the magnetic field and temperature protocol used throughout this work. The upper panel shows the temperature evolution, including zero-field cooling (black) and field-cooling segments (red), with measurement points labeled 1–5. The lower panel shows the corresponding applied magnetic field sequence as a function of time, with an applied field H=1.5G during the field-cooling segments. **d** Scanning SQUID flux maps acquired at the protocol stages labeled 1–5 in panel **c**, shown on the same color scale. After zero-field cooling (ZFC) **1** and after field removal following ZFC **3**, no flux contrast relative to the background is observed. In contrast, a finite flux signal is present during the in-field scans **2, 4** and persists as a remanent signal after field removal following field cooling **5**. Dashed lines indicate the locations of the line profiles shown in panel **e**. **e** One-dimensional line profiles taken across the flake along the dashed lines shown in panel **d**. The blue curve corresponds to the in-field response after zero-field cooling **2**, the green curve to the in-field response after field cooling **4**, and the dashed orange curve to the remanent flux after field removal following field cooling **4, 5**. The inset highlights the difference between the profiles, illustrating the magnitude of the remanent signal. Scale bars in all panels correspond to 10 µm.

To establish the existence of magnetic memory in the absence of vortices, we implement the field and temperature protocol shown in Fig. 1c on thin 4Hb-$TaS_2$ flakes. Figure 1a shows an optical image of a representative flake, and the accompanying susceptometry map (Fig. 1b) locates the flake prior to flux imaging. After zero-field cooling (ZFC) from above $T_*$ through $T_c$ to base temperature, the flake is indistinguishable from the substrate in flux at zero applied field (Fig. 1d1), indicating the absence of any detectable spontaneous signal within the measurement noise. Upon applying a small magnetic field (H = 1500 mG), a finite diamagnetic flux contrast appears (Fig. 1d2), and once the field is removed, no remanent flux signal is observed (Fig. 1d3). In contrast, when the flake is trained by field cooling through the same temperature range, the in-field flux contrast is enhanced (Fig. 1d4) and a finite signal persists as a remanent signal after the field is removed (Fig. 1d5). This remanent flux signal constitutes the measurable manifestation of magnetic memory in the screening currents.

Note that the remanent flux measured on the flake (Fig. 1d5) points in a direction opposite to the training field (which is now off), as opposed to the direction of the spontaneous vortices measured in Ref.[20], whose polarity aligns with the direction of the training field. To quantify the remanent flux signal, we extract one-dimensional line profiles taken perpendicular to the flake edge across the full width of the sample, as illustrated by the dashed line in Fig. 1d and the corresponding profiles in Fig. 1e. We define the signal amplitude $\Delta\Phi$ as the peak-to-peak contrast of each profile.

The in-field amplitude after field cooling through $T_*$, $\Delta\Phi_{FC}$, is approximately 12% larger than the corresponding response obtained after zero-field cooling, $\Delta\Phi_{ZFC}$ (Fig. 1e). After the applied field is removed, the remanent amplitude $\Delta\Phi_{rem}$ is likewise approximately 12% of $\Delta\Phi_{ZFC}$. Repeating this analysis for multiple line profiles across a given flake yields $\Delta\Phi_{rem}/\Delta\Phi_{ZFC} \approx 0.12 \pm 0.002$ (Supplementary Fig. S1). Across five flakes, each measured under its respective training field, we obtain an average value $\Delta\Phi_{rem}/\Delta\Phi_{ZFC} \approx 0.11 \pm 0.01$. These results are consistent with the spontaneous vortex density observed in Ref.[20]

Beyond the magnitude of the response, its spatial structure exhibits an astonishing simplicity. Figure 1e shows that the remanent profile measured at zero applied field closely follows the shape of the screening profile generated by an external field after zero-field cooling. To compare the two directly, we subtract the remanent profile from the in-field profile measured after field cooling and compare the result with the zero-field-cooled in-field response. The two profiles coincide over the entire scan, demonstrating that the field-cooled response can be decomposed as

$$\Delta\Phi_{FC} = \Delta\Phi_{ZFC} + \Delta\Phi_{rem}. \quad (1)$$

Equivalently, the remanent signal has the same spatial profile as the screening response to an externally applied field, differing only by an overall scale factor of approximately 0.12. Thus, the magnetic memory does not generate a distinct flux pattern but instead appears as a proportional modification of the Meissner response. We further confirm that the memory is stable in time (Supplementary Fig. S3) and that by increasing the flake size (see Supplementary Fig. S4), we revert to the previously reported vortex memory (Ref.[20]). As we show below, this observation places strong constraints on the possible origin of the memory.

## Identification of a distinct intermediate regime between $T_c$ and $T_*$

To probe the history dependence of the magnetic memory itself, we implement the protocol shown in Fig. S6. The flake is first trained by field cooling through $T_*$ and $T_c$ in an applied field H, after which the field is removed, and a flux scan is acquired. The temperature is then raised to a fixed cycling temperature $T_{cycle}$, with $T_c < T_{cycle} < T_*$, in the absence of an applied field, and subsequently field-cooled back below $T_c$ under a new training field. The field is again removed, and the remanent response is remeasured. Repeating this sequence for a series of positive and negative training fields generates a closed thermoremanent magnetization (TRM) loop of $\Delta\Phi_{rem}$ versus the training field H, shown in Fig. 2b for multiple $T_{cycle}$ (see also Fig. S6 for $T_{cycle}$ = 3.5 K).

To compare TRM loops obtained from samples with different geometries, including bulk samples, the remanent response must be expressed in a geometry-independent form. In raw flux units, identical training fields produce different amplitudes $\Delta\Phi_{rem}$ across samples and even across scan positions within a single flake, reflecting variations in geometric coupling to the SQUID pickup loop. We therefore convert $\Delta\Phi_{rem}$ into a field equivalent using an empirical, flake-specific calibration obtained at base temperature (2 K) from the slope of the zero-field-cooled Meissner response $\Delta\Phi_{ZFC}$ as a function of applied field (see Supplementary Section S5). This mapping assigns to each measured $\Delta\Phi_{rem}$ the external field that would generate the same screening response after zero-field cooling, providing a common and geometry-independent measure of memory strength.

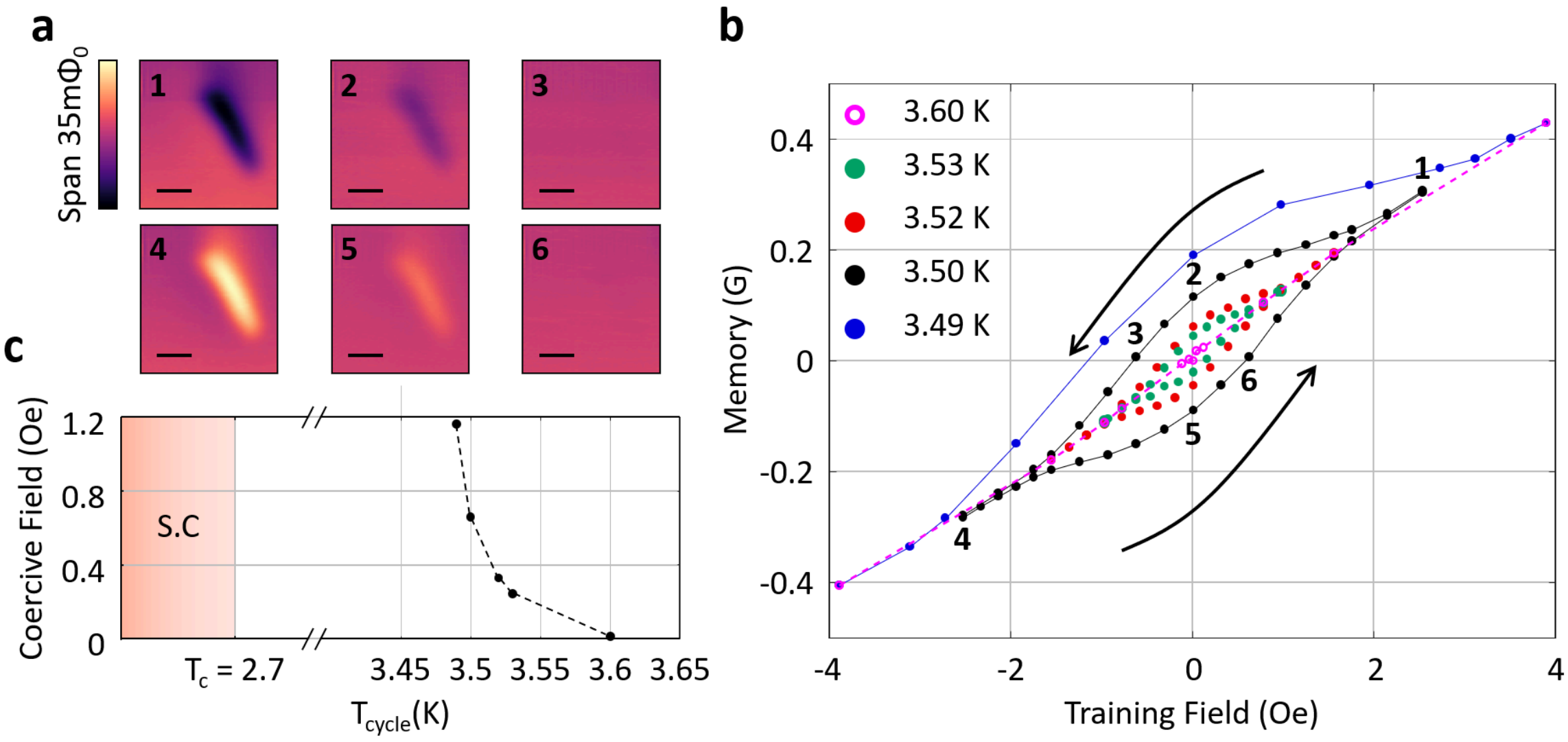

**Fig. 2 | Temperature evolution of field-dependent TRM loops probed through remanent flux. a** Scanning SQUID flux maps acquired at selected points along the TRM loop corresponding to the numbered points indicated in panel **b**. All maps are shown on the same color scale with the scale bar 10um. **b** Remanent flux, expressed as an equivalent magnetic field, plotted as a function of training field H for several cycling temperatures $T_{cycle}$. At low $T_{cycle}$ the memory exhibits a clear hysteresis-like loop with finite coercivity, while with increasing $T_{cycle}$ the loop area decreases continuously and collapses into a single-valued linear relation at $T_*$. Arrows indicate the sweep direction. **c** Coercive field H*, extracted from the zero crossing of the remanent memory during field sweeps, as a function of cycling temperature $T_{cycle}$. H* decreases monotonically with increasing temperature and approaches zero at $T_*$, marking the upper boundary of the intermediate regime in which hysteretic-like behavior in the memory is observed.

The resulting loops are plotted in Fig. 2b as field equivalent versus training field for several representative values of $T_{cycle}$. Unlike a conventional magnetic hysteresis loop, every point is measured at zero applied field; the horizontal axis denotes the field used to prepare the state rather than the field present during measurement. The data reveal two distinct contributions to the memory. First, there is an approximately linear dependence of the remanent memory on the training field, corresponding to an equivalent memory field that is roughly 10% of the applied training field. Superimposed on this linear response is a hysteretic component that depends on the history of the training field sweep. In this sense, the system exhibits a form of "hysteresis within hysteresis": the remanent memory retains a persisting history dependence despite every measurement being performed at zero applied field. As $T_{cycle}$ is increased toward $T_*$, this hysteretic component progressively shrinks while the underlying linear dependence remains.

To quantify the evolution of the hysteretic contribution, we extract the coercive field $H_*$, defined as the training field at which the remanent memory crosses zero during a field sweep. The resulting $H_*(T)$ is shown in Fig. 2c. The coercive field decreases monotonically with increasing temperature and approaches zero as T approaches $T_*$. Importantly, this marks the loss of multistability in the memory response, leaving only the reversible component that depends linearly on the training field.

The loss of coercivity at $T_*$ identifies the characteristic temperature associated with the underlying time-reversal-symmetry-breaking phase above $T_c$. The disappearance of hysteresis at $T_*$ demonstrates that the multistable phase responsible for the memory is not tied to superconducting coherence. The presence of thermoremanent loops and finite coercivity below $T_*$ further shows that this phase possesses multiple metastable configurations that can be selected by an external magnetic field. These observations are consistent with the previously proposed TRSB scenario for 4Hb-TaS2, in which distinct field-trainable states emerge above the superconducting transition and persist independently of superconductivity itself.

## Discussion

The central experimental observation is that the remanent response is a scaled replica of the ordinary Meissner response. As shown in Fig. 1, the trained response obeys the empirical relation Eq. (1), while $\Delta\Phi_{rem}$ exhibits the same spatial profile as $\Delta\Phi_{ZFC}$ up to a scale factor. Any microscopic description of the memory must therefore explain why the remanent state modifies the superconducting screening response without generating an independent magnetic-field profile.

As discussed above, one class of interpretations associates the remanent state with weak intrinsic magnetization that is stabilized or amplified by superconductivity, spin–orbit coupling, or electronic correlations. Proposed mechanisms of this type include weak magnetic order coupled to unconventional superconductivity, spin-glass–like freezing, or related altermagnetic scenarios[26–28]. In all such pictures, the assumed Ginzburg-Landau energy density functional is taken to be

$$f - f_{\rho} = \frac{\rho_s}{2}(\nabla\theta - eA)^2 + \frac{1}{2\mu_0}[\nabla \times A - \mu_0(H + M)]^2 \ldots ,$$

where $M$ is the magnetization, which is enhanced below $T_c$, $\Psi = \sqrt{\rho_s}\exp(i\theta)$ is the order parameter, and $f_{\rho}$ includes the non-field dependent terms (see SM). The magnetic part may originate from magnetic degrees of freedom, non-unitary, or chiral superconductivity. In other words, the remanent state is a response to a real magnetic field distribution generated by a physical magnetization within the sample. This family of mechanisms can be directly addressed because our measurements provide spatially resolved flux profiles. To understand this, we calculate the magnetic field and flux distribution expected from a weak static magnetization. (The full derivation and numerical implementation of this calculation are provided in Supplementary Note S7.) The key distinction between an externally applied field and a field generated by an internal magnetization lies in the spatial location of the source currents. An external field is generated by currents outside the sample and subsequently screened by Meissner currents. By contrast, an internal magnetization generates bound currents $J_M = \nabla \times M$ within the sample itself, requiring the screening currents to redistribute a field whose sources are internal to the superconductor. As a result, magnetization-induced fields generally produce spatial profiles that are qualitatively distinct from those arising from external fields.

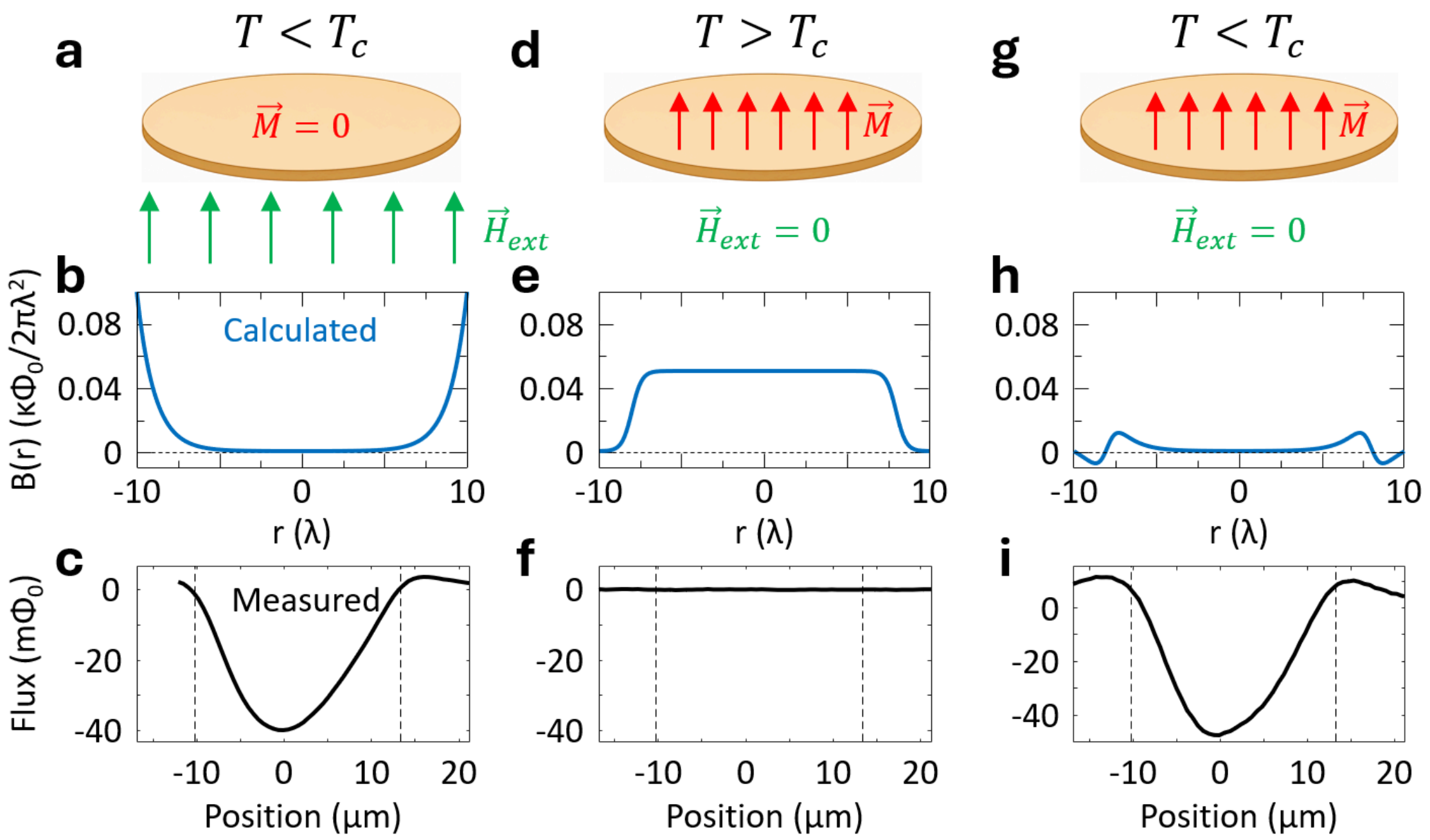

**Fig. 3 | Calculated and measured flux profiles for three distinct scenarios. a-c** Superconducting disk below $T_c$ with no intrinsic magnetization in an applied external field: **a** schematic, **b** calculated field profile showing boundary-localized Meissner screening, and **c** the experimentally measured profile in a corresponding configuration, with an external field of H = 0.2G. The calculation reproduces the measured zero-field-cooled in-field response, with small deviations arising from geometric factors and measurement resolution. The dashed lines in **c**, **f** and **i** represent the sample edges of the scanned flake. **d-f** Magnetized disk above $T_c$ with no applied field: **d** schematic, **e** calculated profile, and **f** measured profile after training with H = 2G. **g-i** Magnetized disk below $T_c$ with no applied field: **g** schematic, **h** calculated profile, and **i** measured profile after training with H = 2G. The qualitative discrepancy for this case is clear. It underscores that the measured profile, which is found for a configuration with zero external field (**d, g**) but matches the expected profile for a nonzero external field, is not compatible with the interpretation of an internal uniform magnetization.

Fig. 3 compares the experimental measurement to the expected profile for two distinct scenarios within the same disk geometry. The first scenario (1st column) is a superconducting disk below $T_c$ in the absence of intrinsic magnetization and in a small external magnetic field. In this case, the calculated screening currents at the edge, which decay over a length scale **λ**, cancel the magnetic field in the interior of the sample. The calculated profile closely reproduces the measured flake after cooling it with no external field active through $T_*$, or ZFC. After ZFC, we applied an external magnetic field and measured the flake. The schematic configuration, calculated field distribution, and measured line profile are shown in Fig. 3(a–c). The differences between the calculated and measured profiles are attributable to geometric factors and the measurement response, including the sensor's point spread function and its height above the flake, and do not affect the qualitative agreement. The second scenario (2nd and 3rd columns) describes the case of a polarized magnetic disk with no applied external field, both below and above $T_c$ [Fig.~3(d--f) and Fig.~3(g--i)]. In magnetization-based scenarios invoking amplified intrinsic magnetic order, these configurations correspond to the remanent trained state. However, the calculated flux distributions are qualitatively inconsistent with the measured remanent profile. The key discrepancy is the large interior signal in the experiment (as compared with the region outside the sample). In contrast, the measured remanent signal shows a clear and finite flux throughout the interior of the sample. Moreover, the flux cross-section is identical to the one observed in an untrained sample in a corresponding magnetic field (Supp. Fig. S2).

This difference reflects a fundamental distinction between fields generated by internal magnetization currents and those arising from superconducting screening currents. In the former, both the source and screening currents reside within the sample, leading to substantial cancellation of the internal field. In the latter, screening currents redistribute an externally imposed or effective field, producing the boundary-localized structure characteristic of Meissner screening. The observed remanent response follows the latter behavior and is therefore incompatible with a description in terms of static internal magnetization. Ruling out finite-thickness effects as the origin of this discrepancy. Following Pearl[29], the London response of a thin slab of width $d$ is given by $A(q_{\|}) = \alpha\, d\, j_M/(2|q_{\|}| + 1/\lambda_p)$, where $q_{\|}$ is the in-plane wavevector and $\lambda_p \approx 2.5\mu m$, is the Pearl length. In our samples, the lateral dimensions are much larger than $\lambda_p$, placing the system in the three-dimensional (albeit thin film) limit. In this regime, finite-thickness corrections do not qualitatively alter the spatial structure of the response and therefore cannot reconcile the magnetization-based predictions with the observed interior flux. We also account for the finite distance between the SQUID and the sample (see Supplementary materials).

Another inconsistency between the measurements and the internal magnetization scenario is the direction of the measured flux. In the previous experiment, performed on bulk samples, the flux of the spontaneous vortices appeared in the **same direction as the training field**. In the current experiment, performed on flakes, the measured remanent flux **opposes the direction of the training field**. Thus, the flux pattern in both bulk and flake samples is what one would expect Meissner shielding currents to generate in response to an external field, which is absent.

Taken together, the spatial profile and the additivity relation, Eq. (1), are difficult to reconcile with a static internal magnetization. Instead, they suggest that the superconducting condensate responds to an additional effective field that is not directly detected by the SQUID. This naturally accounts for the proportionality between the remanent and ordinary Meissner responses: both correspond to screening of a field experienced by the condensate, while only the electromagnetic component is directly measured.

One possible realization is an emergent gauge field associated with hidden electronic degrees of freedom, such that ($A \rightarrow A+a$), where ($a$) denotes an internal gauge field. Related proposals for 4Hb-$TaS_2$ invoke fractionalized or topologically nontrivial states, including vison–vortex transmutation in a Z2 gauge structure, chiral spin-liquid physics, or effective gauge fields arising from spin–orbit coupling, time-reversal-symmetry breaking, and charge inhomogeneity[27,30–32]. In such scenarios, the superconducting condensate responds to the combined external and internal gauge fields, whereas the SQUID detects only the electromagnetic field. For this picture to account for magnetic training, the internal degree of freedom must also couple, directly or indirectly, to the external magnetic field, allowing it to be trained by crossing $T_*$.

The temperature dependence provides an additional constraint on this hidden state. The disappearance of coercivity at $T_* > T_c$ shows that the memory is established independently of superconducting coherence, even though its remanent response is read out through Meissner screening only below $T_c$. Together with the persistence of memory in the vortex-free regime, this identifies the effect as a property of an electronic phase preceding superconductivity rather than of trapped superconducting flux.

Our measurements therefore establish vortex-free Meissner memory as a probe of the hidden electronic state in 4Hb-$TaS_2$. Its spatial form strongly disfavors a simple static-magnetization origin and is consistent with an effective field acting on the superconducting condensate. Whether this field originates from an emergent gauge structure or another time-reversal-symmetry-breaking electronic order remains an open microscopic question.

# Methods

**Sample preparation**

Single crystals of 4Hb-$TaS_{1.99}Se_{0.01}$ were grown using the chemical vapor transport (CVT) method. A stoichiometric mixture of tantalum (Ta), sulfur (S) and selenium (Se) was sealed in a quartz ampoule under vacuum. The mixture underwent a solid-state reaction, forming a boule of 4Hb-$TaS_{1.99}Se_{0.01}$. The boule was crushed and placed in a 200-mm-long quartz ampoule with 16-mm diameter. Iodine was added as a transport agent, and the ampoule was sealed under vacuum. The ampoule was then placed in a three-zone furnace, where the hot ends were heated to 800$^{o}$ C and the middle part was kept at 750$^{o}$ C. After about 30 days, the ampoule was quenched in cold water. Crystals having a few mm in size are found in the cold part of the ampoule. 4Hb-$TaS_{1.99}Se_{0.01}$ flakes were exfoliated using the blue-tape technique and immediately transferred onto cleaned SiOx/Si substrates comprising a 285-nm-thick SiOx layer on p-type Si.

**Scanning SQUID microscopy**

Simultaneous local magnetometry and susceptometry measurements were performed by scanning a planar SQUID with a 0.75-μm pick-up loop over a cleaved surface of the sample[24,25]. The cleaved sample was glued onto an aluminium-clad printed circuit board sample holder using GE Varnish (Lakeshore Cryotronics). Susceptometry measurements were performed by a mutual inductance measurement using a second coil (diameter 3 μm) concentric to the SQUID's pick-up loop. An alternating current (1 mA RMS, at about 1.1 kHz) applied to this coil generates a local magnetic field of about 1.5 G RMS. The superconductor screens this field, generating a change in the flux penetrating the SQUID's pick-up loop. This change was recorded simultaneously to the d.c. measurement using a lock-in amplifier. Scans were performed at a constant sensor–sample distance. An external magnetic field was applied using a copper coil wrapped around the microscope set-up. Earth's magnetic field was reduced by including two concentric low-temperature μ-metal (A4K, Amuneal) cylindrical shields with diameters of 96 mm and 100 mm around the microscope and coil, achieving fields as small as 5–10 mG near the sample. Further cancellation was achieved by applying a small current through the copper coil, reaching fields of about 3 mG. 'Zero field' measurements were performed under these conditions (about 3 mG). Temperature was monitored with a silicon diode and thermal cycles were performed at a rate of about 0.5 K min−1 with an open-loop control. The peak temperature at each cycle was held for 3 min before the sample was cooled. Before thermal cycles, the SQUID was retracted over 100 μm from the sample to avoid field distortions owing to the superconducting material on the sensor.

# Acknowledgements

We thank H. Beidenkopf, P. Lee, J. Venderbos and E. Berg for useful discussions. S. Gur, T.R. Devidas, E. Persky, R. A. Gofman and B. Kalisky were supported by the European Research Council Grant No. ERC-2019- OG-866236, the Israeli Science Foundation Grant No. ISF-228/22. A. Klein acknowledges the support of the Israeli Science Foundation grant no. 3152/25. JR acknowledges funding by the Simons Foundation under the collaborative grant “New Frontiers in Superconductivity". SKS acknowledges the support of the Israeli Science Foundation grant no. 915/24. The work at the Technion was supported by the Israeli Science Foundation under grant number ISF-568/25. This work was supported by the German–Israeli Project Cooperation (DIP), Grant Nos. 529677299 and 3970/1-1.

## Data availability

The data that support the findings of this study are available from the corresponding authors upon reasonable request.

## Author contributions

A. Kanigel and B. Kalisky initiated this study. S. Gur, T.R. Devidas and B. Kalisky designed the experiments and performed the scanning SQUID measurements. E.Persky performed preliminary scanning SQUID measurements. I. Feldman, A. Almoalem and A. Kanigel prepared the sample. A. Almoalem performed the global characterization measurements. S.K. Saha, A. Klein, M.H. Fischer, J. Ruhman performed the theoretical calculations. S. Gur, B. Kalisky, A. Kanigel, S.K. Saha, A. Klein, M.H. Fischer and J. Ruhman discussed the data and interpreted the results. All authors contributed to writing the manuscript.

## Additional information

### Supplementary information

The online version contains supplementary material available at <link>

### Correspondence and requests for materials

Correspondence and requests for materials should be addressed to B. Kalisky beena@biu.ac.il.

---

1. Brandt, E. H. The flux-line lattice in superconductors. *Rep. Prog. Phys.* **58**, 1465–1594 (1995).
2. Bean, C. P. Magnetization of high-field superconductors. *Rev. Mod. Phys.* **36**, 31–39 (1964).
3. Luke, G. M. *et al.* Time-reversal symmetry-breaking superconductivity in $Sr_2RuO_4$. *Nature* **394**, 558–561 (1998).
4. Hillier, A. D., Quintanilla, J. & Cywinski, R. Evidence for time-reversal symmetry breaking in the noncentrosymmetric superconductor $LaNiC_2$. *Phys Rev Lett* **102**, 117007 (2009).
5. Singh, R. P. *et al.* Detection of time-reversal symmetry breaking in the noncentrosymmetric superconductor Re6Zr using muon-spin spectroscopy. *Phys Rev Lett* **112**, 107002 (2014).
6. Xia, J., Maeno, Y., Beyersdorf, P. T., Fejer, M. M. & Kapitulnik, A. High resolution polar Kerr effect measurements of $Sr_2RuO_4$: evidence for broken time-reversal symmetry in the superconducting state. *Phys Rev Lett* **97**, 167002 (2006).

7. Sigrist, M. & Ueda, K. Phenomenological theory of unconventional superconductivity. *Rev. Mod. Phys.* **63**, 239–311 (1991).
8. Fernandes, R. M., Orth, P. P. & Schmalian, J. Intertwined Vestigial Order in Quantum Materials: Nematicity and Beyond. *Annu. Rev. Condens. Matter Phys.* **10**, 133–154 (2019).
9. Motrunich, O. I. Orbital magnetic field effects in spin liquid with spinon Fermi sea: Possible application to κ-$(ET)_2Cu_2(CN)_3$. *Phys. Rev. B Condens. Matter* **73**, 155115 (2006).
10. Sodemann, I., Chowdhury, D. & Senthil, T. Quantum oscillations in insulators with neutral Fermi surfaces. *Phys. Rev. B.* **97**, (2018).
11. Senthil, T. & Fisher, M. P. A. $Z_2$ gauge theory of electron fractionalization in strongly correlated systems. *Phys. Rev. B Condens. Matter* **62**, 7850–7881 (2000).
12. Wilson, J. A., Di Salvo, F. J. & Mahajan, S. Charge-density waves and superlattices in the metallic layered transition metal dichalcogenides. *Adv. Phys.* **24**, 117–201 (1975).
13. Law, K. T. & Lee, P. A. 1T-$TaS_2$ as a quantum spin liquid. *Proceedings of the National Academy of Sciences* **114**, 6996–7000 (2017).
14. He, W.-Y., Xu, X. Y., Chen, G., Law, K. T. & Lee, P. A. Spinon Fermi Surface in a Cluster Mott Insulator Model on a Triangular Lattice and Possible Application to 1T-$TaS_2$. *Phys. Rev. Lett.* **121**, 046401 (2018).
15. Nagata, S. *et al.* Superconductivity in the layered compound 2H-$TaS_2$. *J. Phys. Chem. Solids* **53**, 1259–1263 (1992).
16. Ribak, A. *et al.* Chiral superconductivity in the alternate stacking compound 4Hb-$TaS_2$. *Science Advances* **6**, eaax9480 (2020).
17. Nayak, A. K. *et al.* Evidence of topological boundary modes with topological nodal-point superconductivity. *Nat. Phys.* **17**, 1413–1419 (2021).
18. Silber, I. *et al.* Two-component nematic superconductivity in 4Hb-$TaS_2$. *Nat. Commun.* **15**, 1–6 (2024).
19. Almoalem, A. *et al.* The observation of π-shifts in the Little-Parks effect in 4Hb-$TaS_2$. *Nat. Commun.* **15**, 4623 (2024).
20. Persky, E. *et al.* Magnetic memory and spontaneous vortices in a van der Waals superconductor. *Nature* **607**, 692–696 (2022).
21. Auslaender, O. M. *et al.* Mechanics of individual isolated vortices in a cuprate superconductor. *Nat. Phys.* **5**, 35–39 (2009).
22. Tonomura, A. *et al.* Observation of individual vortices trapped along columnar defects in high-temperature superconductors. *Nature* **412**, 620–622 (2001).
23. Moshchalkov, V. V. *et al.* Pinning by an antidot lattice: The problem of the optimum antidot size. *Phys. Rev. B Condens. Matter* **57**, 3615–3622 (1998).
24. Gardner, B. W. *et al.* Scanning superconducting quantum interference device susceptometry. *Rev. Sci. Instrum.* **72**, 2361–2364 (2001).
25. Huber, M. E. *et al.* Gradiometric micro-SQUID susceptometer for scanning measurements of mesoscopic samples. *Rev. Sci. Instrum.* **79**, 053704 (2008).
26. Liu, C., Chatterjee, S., Scaffidi, T., Berg, E. & Altman, E. Magnetization amplification in the interlayer pairing superconductor 4Hb−$TaS_2$. *Phys. Rev. B.* **110**, (2024).
27. Levitan, B. A., Oreg, Y. & Berg, E. Anomalous currents and spontaneous vortices in spin-orbit coupled superconductors. *NPJ Quantum Mater* **10**, 58 (2025).
28. Ramires, A. From pure to mixed: Altermagnets as intrinsic symmetry-breaking indicators. *Phys. Rev. Res.* **8**, (2026).
29. Pearl, J. Current distribution in superconducting films carrying quantized fluxoids. *Appl. Phys. Lett.* **5**, 65–66 (1964).
30. König, E. J. Type-II heavy Fermi liquids and the magnetic memory of 4Hb−$TaS_2$. *Phys. Rev. Res.* **6**, (2024).
31. Lin, S.-Z. Kondo enabled transmutation between spinons and superconducting vortices: Origin of magnetic memory in 4Hb−$TaS_2$. *Phys. Rev. Res.* **6**, (2024).
32. Luo, R. L. & Chen, G. V. Is spontaneous vortex generation in superconducting 4Hb-$TaS_2$ from vison-vortex nucleation with $Z_2$ topological order? *Quantum Front.* **3**, (2024).

# Supplementary Material for
# Evidence for an effective gauge field in flakes of superconducting 4Hb-TaS2

Shalev Gur[1,2], T.R. Devidas[1,2], Eylon Persky[1,2], Roni Anna Gofman[1,2], Irena Feldman[3], Avior Almoalem[3], Sudip Kumar Saha[1], Mark H. Fischer[4], Avraham Klein[1], Jonathan Ruhman[1], Amit Kanigel[3] & Beena Kalisky[1,2✉]

[1]Department of Physics, Bar Ilan University, Ramat Gan, Israel. [2]Bar Ilan Institute of Nanotechnology and Advanced materials, Bar Ilan University, Ramat Gan, Israel. [3]Department of Physics, Technion–Israel Institute of Technology, Haifa, Israel. [4]Department of Physics, University of Zurich, Z¨urich, Switzerland
✉e–mail: beena@biu.ac.il

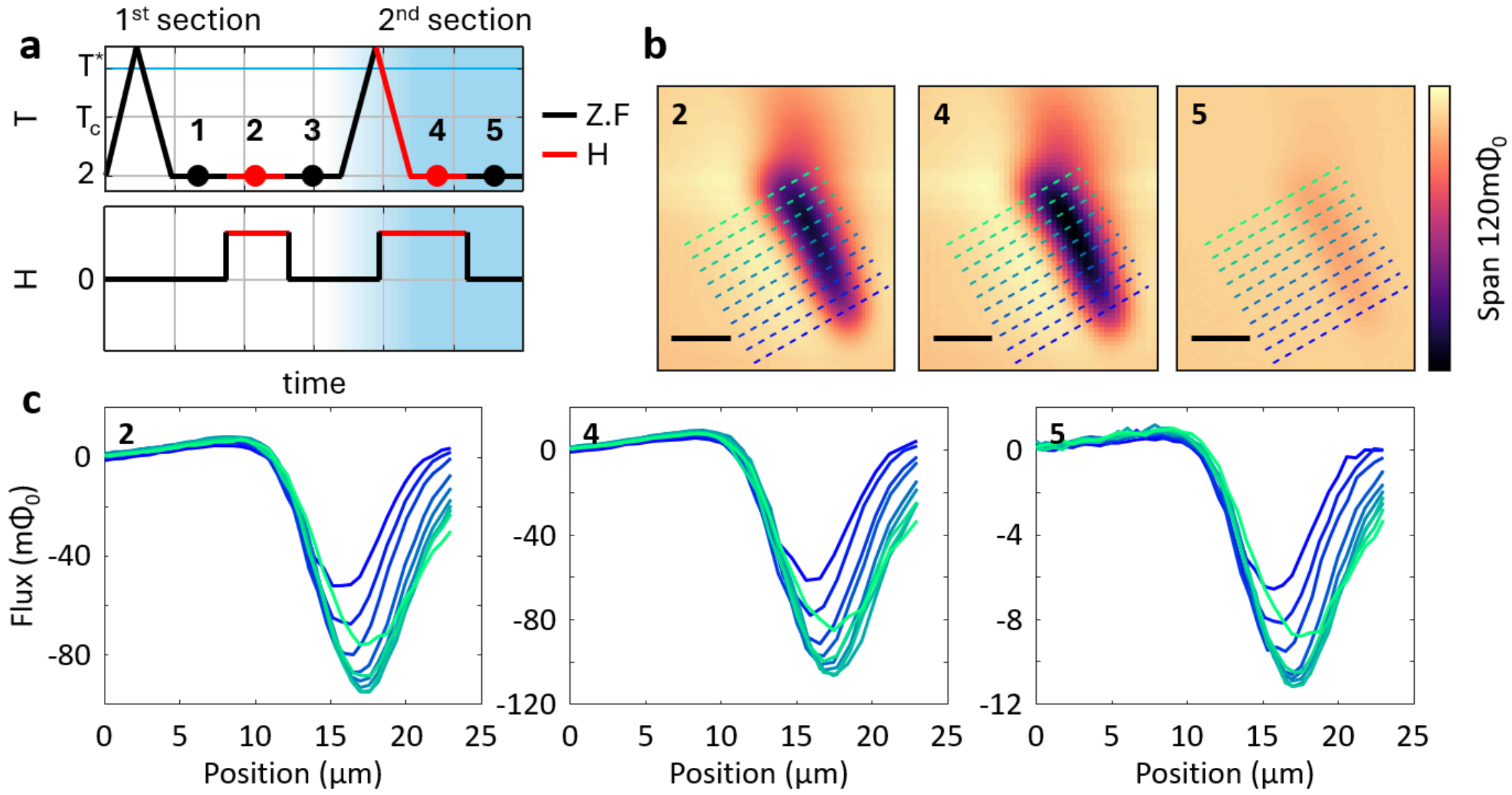

**Supplementary Fig. S1 | Spatial homogeneity of the remanent Meissner signal. a** Schematic of the magnetic field and temperature protocol as seen in the main text. **b** Magnetometry maps of a 50*40μm2 area taken over the flake in the states **2, 4, 5** created by following the experimental procedure **a**. Dashed lines indicate multiple line cuts taken at different positions across the flake. Scale bar corresponds to 10 μm.

**c** One-dimensional line profiles extracted along the dashed lines shown in **b**. By following the methodology shown in the main text, the average memory strength was calculated to be 11.99% with a STD of 0.256%.

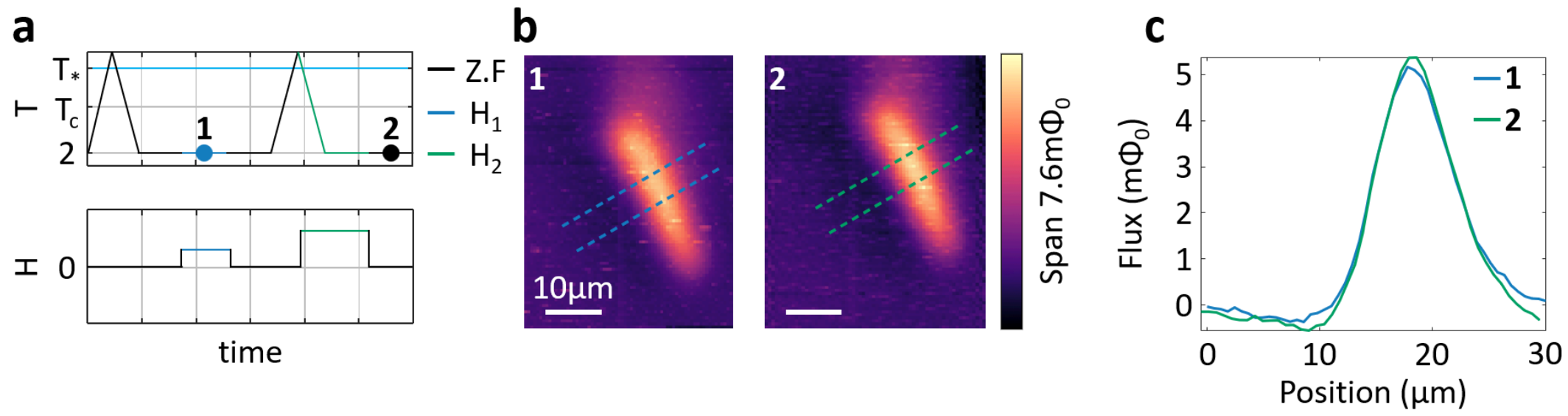

**Supplementary Fig. S2 | Validation of spatial uniformity via comparison to externally applied fields. a** Schematic of the protocol used to compare the spatial structure of the memory-induced magnetic response with that generated by an externally applied magnetic field. Two states are prepared using magnetic fields H1 = 78 mG and H2 = 780 mG. Owing to the approximately order-of-magnitude difference between the screening response generated by the memory effect and that generated directly by an applied field, these field values are chosen such that the resulting signal amplitudes are comparable. **b** Scanning SQUID flux maps of the same flake in the two states defined in panel **a**. The left panel shows the response to the externally applied field H1, and the right panel shows the remanent response generated by the memory effect following application of H2. Both maps are shown on the same color scale. **c** One-dimensional line profiles of the flux signal for the two states shown in panel **b**, obtained by averaging over several line profiles taken parallel to the dashed lines indicated in panel **b**. The profile corresponding to H1 is shown in blue, and the profile corresponding to H2 is shown in green. The profiles exhibit similar shapes and amplitudes, indicating that the memory-induced magnetic field is spatially uniform and comparable in structure to that produced by an external field.

---

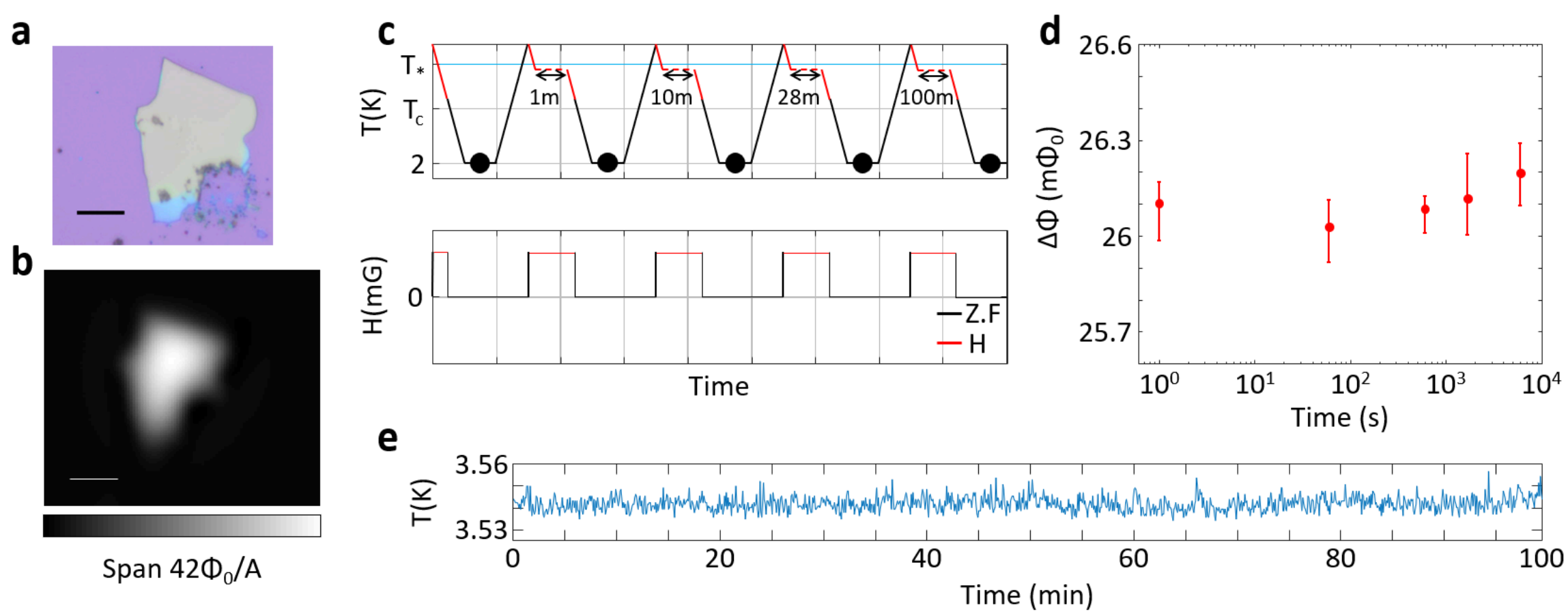

**Supplementary Fig. S3 | Waiting-time independence of the remanent memory. a** Optical image of a representative 4Hb-$TaS_2$ flake used for waiting-time measurements, H = 1G. **b** Scanning SQUID susceptibility image of the same flake, used to identify the flake boundaries and sensor position. **c** Schematic of the temperature and magnetic-field protocol used to probe the dependence of the remanent memory on waiting time. After field cooling to a temperature below $T_*$, the sample is held for a waiting time t before further cooling below $T_c$ under an applied magnetic field. Waiting times of 1 min, 10 min, 28 min, and 100 min are examined. **d** Peak-to-peak flux amplitude ΔΦ of the remanent Meissner signal measured as a function of waiting time. Within experimental uncertainty, no systematic dependence of ΔΦ on waiting time is observed. Errorbar shows a range of values generated by scanning 10 times in a row, without changing the state of the experiment, while the marker shows

the average of all of these valuest. **e** Temperature stability during the waiting-time protocol, showing fluctuations much smaller than the temperature scale relevant for the memory effect.

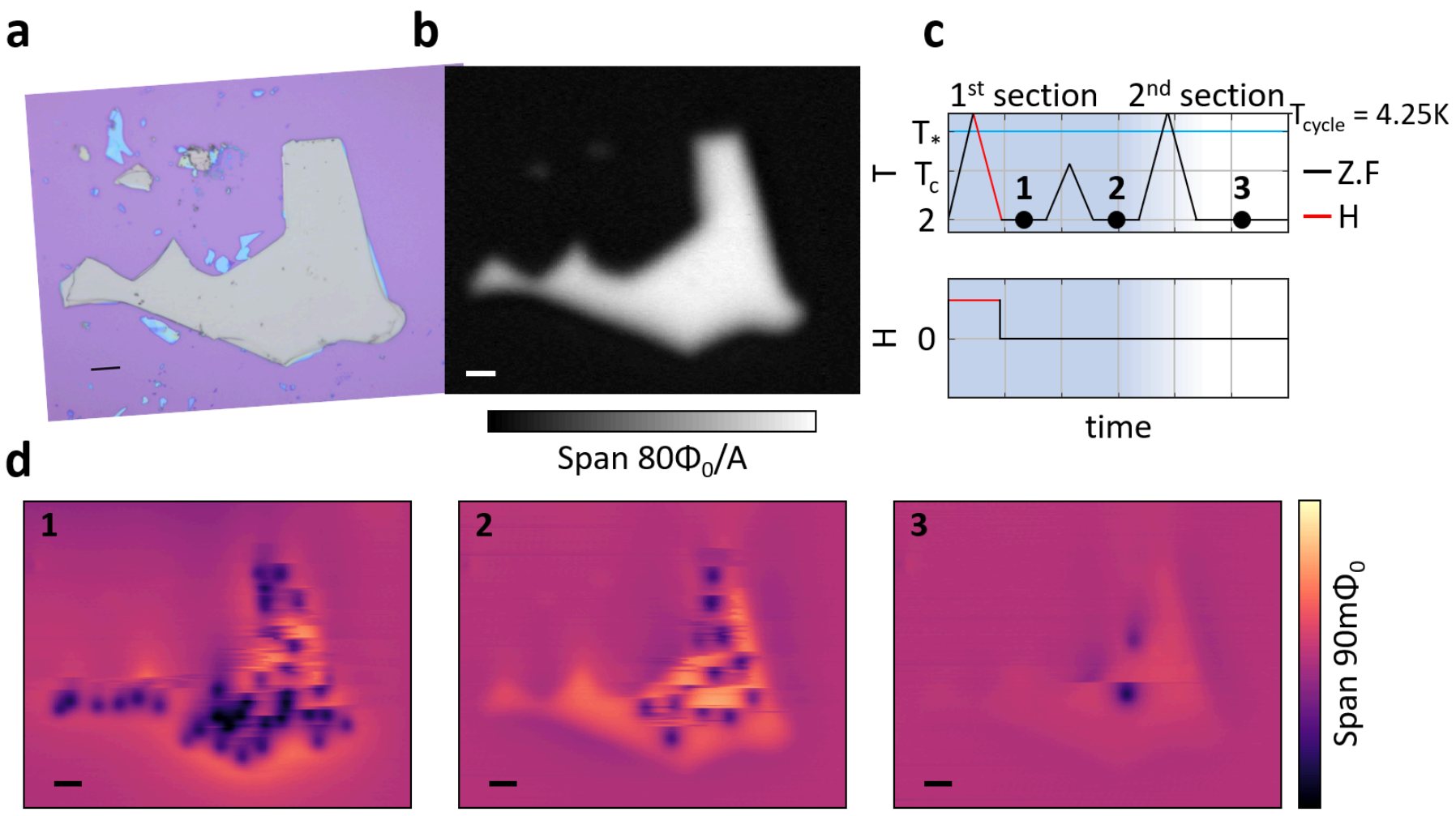

**Supplementary Fig. S4 | Memory behavior in the vortex regime. a** Optical image of a representative 4Hb-$TaS_2$ flake used for vortex-regime measurements. **b** Scanning SQUID susceptibility image of the same flake, used to identify the flake boundaries and sensor position. **c** Schematic of the temperature and magnetic-field protocol used to probe memory in the vortex regime. A magnetic field H = 0.39 G is applied during cooling to access a dilute vortex regime, distinct from the Meissner-regime protocols used in the main text. Although the flakes have similar thickness, the vortex measurements are performed on a substantially larger-area flake; the onset of resolvable vortices is geometry- and pinning-dependent and can occur at different applied fields for different flakes. **d** Scanning SQUID flux maps acquired at the three measurement points indicated in panel **c**. Localized flux features corresponding to individual vortices are observed. The maps reveal 38, 13, and 2 vortices in the three states, respectively. After heating above $T_c$ and cooling under zero field, vortices remain present, whereas vortices disappear only after heating above $T_*$ and cooling under zero field, consistent with the vortex-memory behavior reported previously by Persky *et al.*

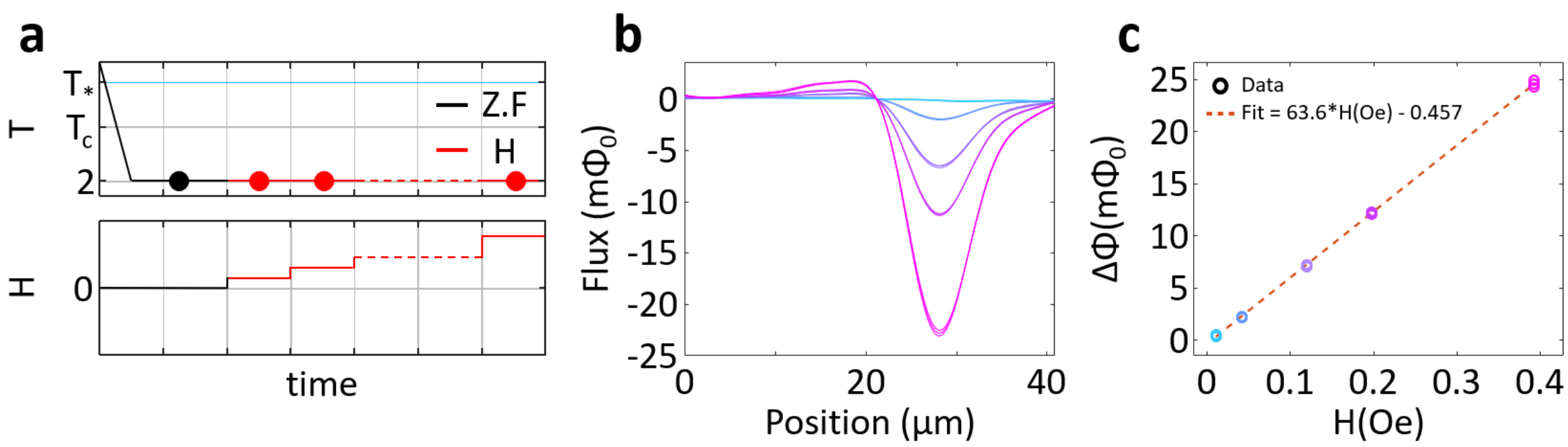

**Supplementary Fig. S5 | Conversion between measured flux and equivalent magnetic field. a** Schematic of the protocol used to calibrate the relationship between the measured scanning SQUID flux signal and the applied magnetic field. The sample is cooled below $T_c$ under zero field, after which

different magnetic fields are applied sequentially at fixed temperature and the resulting flux signal is measured. **b** One-dimensional line profiles of the scanning SQUID flux signal acquired under different applied magnetic fields, extracted along the same spatial cut across the flake. The depth of the flux response increases systematically with increasing applied field while maintaining a similar spatial profile. **c** Peak-to-peak flux amplitude ΔΦ, extracted from the line profiles in panel **b**, plotted as a function of the applied magnetic field H. The data exhibit a linear dependence over the measured field range. The dashed line shows a linear fit, which is used to convert between measured flux (in units of $m\Phi_0$) and an equivalent magnetic field (in units of Oe or G) for the hysteresis plots shown in the main text.

---

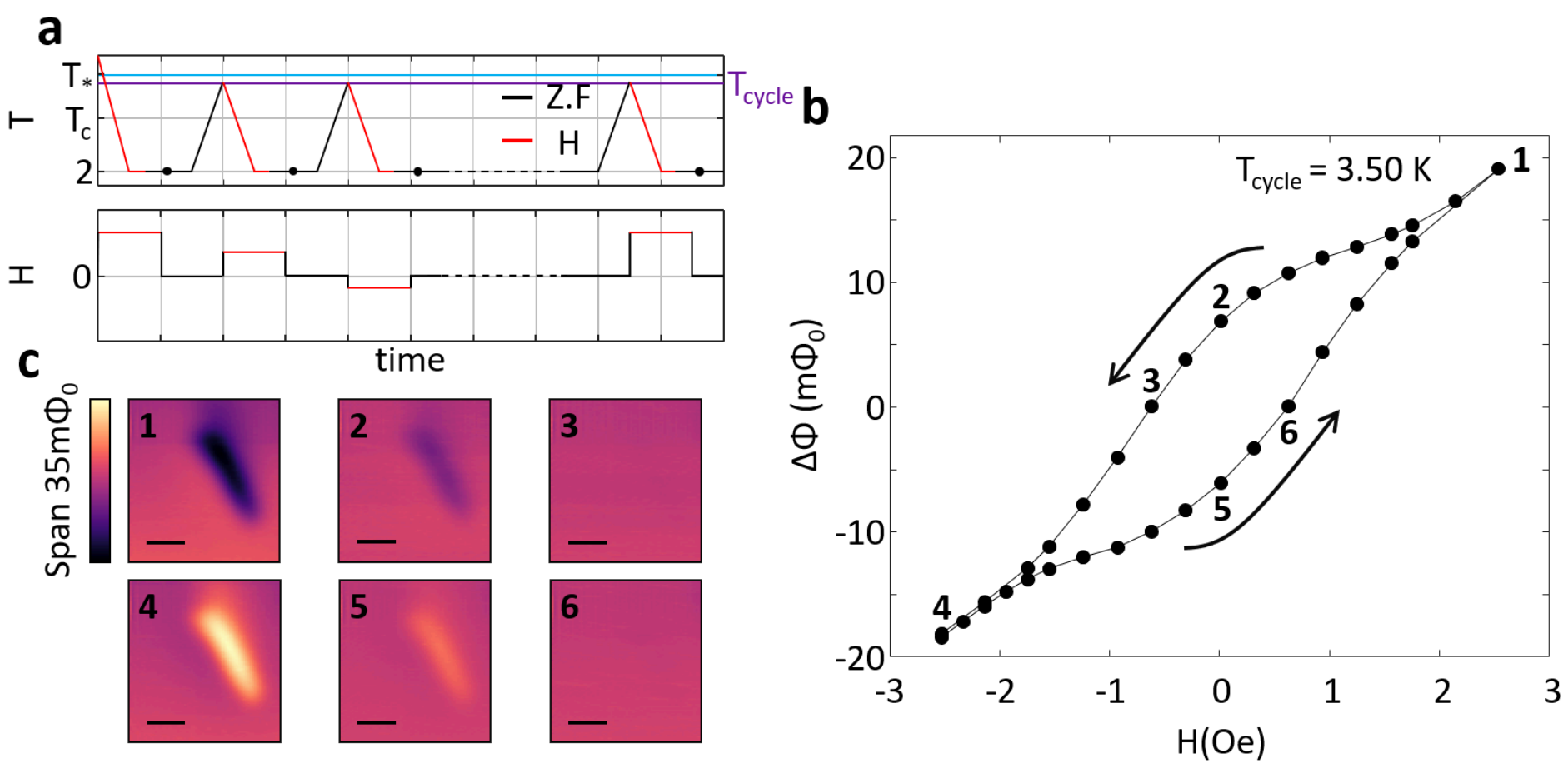

**Supplementary Fig. S6 | Field-dependent TRM probed through remanent screening currents. a** Schematic of the field-dependent TRM protocol used to probe magnetic memory. The upper panel shows the temperature evolution during the protocol, including repeated thermal cycles between a base temperature below $T_c$ and a maximum cycle temperature $T_{cycle}$. Zero-field segments are shown in black and field-applied segments in red. The lower panel shows the corresponding applied magnetic field sequence over time. **b** TRM of the remanent screening currents measured for $T_{cycle}$=3.5K. The remanent signal is characterized by the peak-to-peak amplitude and exhibits a hysteresis-loop-like dependence; we refer to this curve as the TRM loop. **c** Scanning SQUID flux maps acquired at selected points along the TRM loop corresponding to the numbered points indicated in panel **b**. All maps are shown on the same color scale with the scale bar 10um.

---

# Supplementary Section S7 - Theory

## Subsection S7.1 - Solution of the Ginzburg–Landau equations for an infinite superconducting cylinder with finite internal magnetization

In this subsection, we derive the magnetic field and current distribution in the transverse plane of a superconducting cylinder in the presence of a static internal magnetization by solving the Ginzburg-Landau (GL) equations of superconductivity. The free energy functional of the superconducting state in this system can be expressed as

$$F = \mu_0 H_{cm}^2 \int \left[ \frac{1}{2} - |\Psi|^2 + \frac{|\Psi^4|}{2} + \left| \left( \frac{i\vec{\nabla}}{\kappa} + \vec{A} \right) \Psi \right|^2 + \left( \vec{\nabla} \times \vec{A} - \vec{M} \right)^2 \right].$$

(1)

Here, the GL parameter is given by $\kappa = \lambda/\xi$, where $\lambda$ denotes the penetration depth and $\xi$ denotes the coherence length. The spatial distance and the vector potential $\vec{A}$ are measured in units of $\lambda$ and $\hbar/e^*\xi$, respectively. The charge $e^*$ is twice that of the free electron ($e^* = 2e$). The magnetic field $\vec{B} = \vec{\nabla} \times \vec{A}$ and the magnetization $\vec{M}$ are both measured in units of $\hbar\kappa/e^*\lambda^2 = \kappa\phi_0/2\pi\lambda^2$, where $\phi_0 = h/2e$ is the flux quantum. The superconducting current $\vec{j}$ is related to $\vec{A}$ as $\vec{j} = \vec{\nabla} \times \left(\vec{\nabla} \times \vec{A}\right)$, expressed in units of $\hbar\kappa/e^*\mu_0\lambda^3$. The first GL equation is obtained by minimizing Eq. (1) with respect to the order parameter. It assumes the form

$$\left( \frac{i\vec{\nabla}}{\kappa} + \vec{A} \right)^2 \Psi - \Psi + |\Psi|^2 \Psi = 0 .$$

(2)

Similarly, minimizing Eq. (1) with respect to the order parameter $\vec{A}$ leads to the second GL equation, which takes the form

$$- \vec{\nabla} \times \left( \vec{\nabla} \times \vec{A} \right) = \frac{i}{2\kappa} \left( \Psi^* \vec{\nabla}\Psi - \Psi \vec{\nabla}\Psi^* \right) + \vec{A} |\Psi|^2 - \left( \vec{\nabla} \times \vec{M} \right).$$

(3)

We assume the superconducting system to be an infinite cylinder where the field $\vec{B}$ points along the vertical direction ($\vec{e}_z$), and $\vec{A}$ and $\vec{j}$ are along azimuthal ($\vec{e}_\phi$) direction. In this cylindrical symmetry, the order parameter assumes the form $\Psi(r) = f(r)\, exp(-\, iL\phi)$ with winding number $L = 1$. Here, the radius $r$ is measured in units of $\lambda$. We have also introduced the scalar quantity $g = |\vec{A}r\kappa - L|$. Using cylindrical coordinates, the free energy in Eq. (1) assumes the form

$$\frac{F}{\mu_0 H_{cm}^2} = \int 2\pi r\, dr \left[ \frac{1}{2} - f^2 + \frac{f^4}{2} + \frac{1}{\kappa^2} \left( \frac{df}{dr} \right)^2 + \frac{g^2 f^2}{r^2 \kappa^2} + (B - M)^2 \right].$$

(4)

Here, the field is related to $g$ as

$$B = \frac{1}{\kappa r} \frac{dg}{dr}.$$

(5)

The first GL equation in cylindrical coordinates becomes

$$\frac{1}{r} \frac{d}{dr} \left( r \frac{df}{dr} \right) - \frac{f g^2}{r^2} - \kappa^2 f \left( f^2 - 1 \right) = 0 .$$

(6)

The second GL equation can be rewritten in cylindrical coordinates as

$$\frac{d}{dr}\left(\frac{1}{r}\frac{dg}{dr}\right) - \frac{gf^2}{r} - \kappa\frac{dM}{dr} = 0\,.$$

(7)

In this expression, the total current $j$ and the superconducting current $j_{SC}$ have the forms

$$j = -\frac{1}{\kappa}\frac{d}{dr}\left(\frac{1}{r}\frac{dg}{dr}\right) = -\frac{gf^2}{\kappa r} - \frac{dM}{dr}\,.$$

(8)

$$j_{SC} = -\frac{gf^2}{\kappa r}\,.$$

(9)

Finally, we impose the boundary condition for the infinitely long cylinder, with the outer boundary along the radial direction located at $r = r_b$.

$$\frac{df(r)}{dr}\Big|_{r=0} = 0,\ \ g(r=0) = 0,\ \ \frac{df(r)}{dr}\Big|_{r=r_b} = 0,\ \ \frac{dg}{dr}\Big|_{r=r_b} = \kappa r_b B_{ext}\,.$$

(10)

We solve GL equations (Eq. (6) and (7)) via relaxation to the equilibrium of a fictitious time-dependent problem, as shown below.

$$\hat{F} = \frac{1}{r}\frac{d}{dr}\left(r\frac{d}{dr}\right) - \frac{g^2}{r^2} - \kappa^2\left(f^2 - 1\right),\ \ \hat{F} = \frac{\partial f}{\partial t}\,.$$

(11)

$$\hat{G} = \frac{d}{dr}\left(\frac{1}{r}\frac{d}{dr}\right) - \frac{f^2}{r},\ \ \hat{G} = \kappa\frac{dM}{dr}\,.$$

(12)

The steady-state solution of Eqs. (11) and (12) is obtained by iterating in time using a sufficiently small time step (square of the spatial step). The calculation is initialized with $f(r) = 1$ and $g(r) = 0$. At each time step, we have a set of equations from Eq. (11), one for each spatial index, which is solved using the Biconjugate Gradients Stabilized Method to update $f(r)$ for the next time step. The updated $f(r)$ is used in the set of equations from Eq. (12) to compute $g(r)$ for the next time step. The procedure is repeated until the relative change in free energy falls below a small tolerance $\epsilon$.

We solve Eqs. (11) and (12) for an infinitely long superconducting cylinder with the GL parameter $\kappa = 10$. The results for an outer radius $r_b = 30\lambda$ are depicted in Supplementary Fig. S7**.** We consider two cases. In the first case, an external field $B_{ext} = 0.1\,\kappa\phi_0/2\pi\lambda^2$ is applied along $\vec{e}_z$ direction in the absence of any intrinsic magnetization (panel a-c). In the second case, a finite magnetization profile is present, but no external field is applied (panel d-f). The magnetization profiles used in the calculations are shown in the insets of panels (a) and (d). The resulting current and field distributions in the transverse plane of the cylinder are shown in panels (a) and (b), respectively, for the first case and in panels (d) and (e), respectively, for the second case. A detailed discussion of the nature of these profiles, and their inconsistency with the experimental observation is presented in the main paper.

## Subsection S7.2 - Corrections due to the finite distance between the SQUID probe and the sample

In Supplementary Section S7.1, we discussed the derivation of the magnetic field profile in the transverse plane of a superconducting cylinder. However, in the experiment, the SQUID probe is positioned at a finite distance from the sample. In this subsection, we take this into account by computing the field profile at a finite height above the plane.

Let's assume a circular current loop of radius $r$ and finite thickness $L$, centered on the $z$ axis, with its plane perpendicular to $z$ axis. The loop extends from $z =- L/2$ to $z = L/2$, and carries a uniform current density $J(r)$. We want to compute the spatial profile of the resulting magnetic field parallel to the plane of the loop at a certain height $z = h$ above the loop. Considering cylindrical coordinates, the field as a function of radial distance $r'$ from the $z$ axis is given by

$$B_h^{ring}\left(r, r'\right) = \frac{\mu_0 J(r)}{2\pi} \int_{-\frac{L}{2}}^{\frac{L}{2}} \frac{1}{\sqrt{\left(r+r'\right)^2 + (h-z)^2}} \left( K(k) + E(k) \frac{r^2 - r'^2 - (h-z)^2}{\left(r-r'\right)^2 + (h-z)^2} \right) dz \,. \tag{13}$$

Here, $K\left(k_r\right)$ and $E\left(k_r\right)$ are elliptic integrals of the first and second kind with

$$k = \sqrt{\frac{4rr'}{\left(r+r'\right)^2 + \left(h-z\right)^2}} \,. \tag{14}$$

We can think of the transverse plane of a superconducting cylinder as a set of many such loops with radius $- r_b \le r \le r_b$. The contribution of such a set at a height $h$ is given by

$$B_h\left(r'\right) = \sum_{r=-r_b}^{r_b} B_h^{ring}\left(r, r'\right) + B_{ext} \,. \tag{15}$$

We consider $L = 0.1\lambda$ and $h = \lambda$. The superconducting current density for a loop of radius $r$ is given by the current profile $j(r) = j_{SC}(r)$, shown in Supplementary Fig. S7 for the cases with and without static intrinsic magnetization (shown in panels a and d, respectively). The corresponding field profiles $B_h(r)$ for both cases are shown in panels (c) and (f), respectively. However, these profiles are not consistent with the measured remanent flux, shown in Fig. 3 (panels (c), (f), and (i)). Thus, the discrepancy between the theory of static internal magnetization and experimental observation cannot be attributed solely to the finite distance between SQUID and the sample.

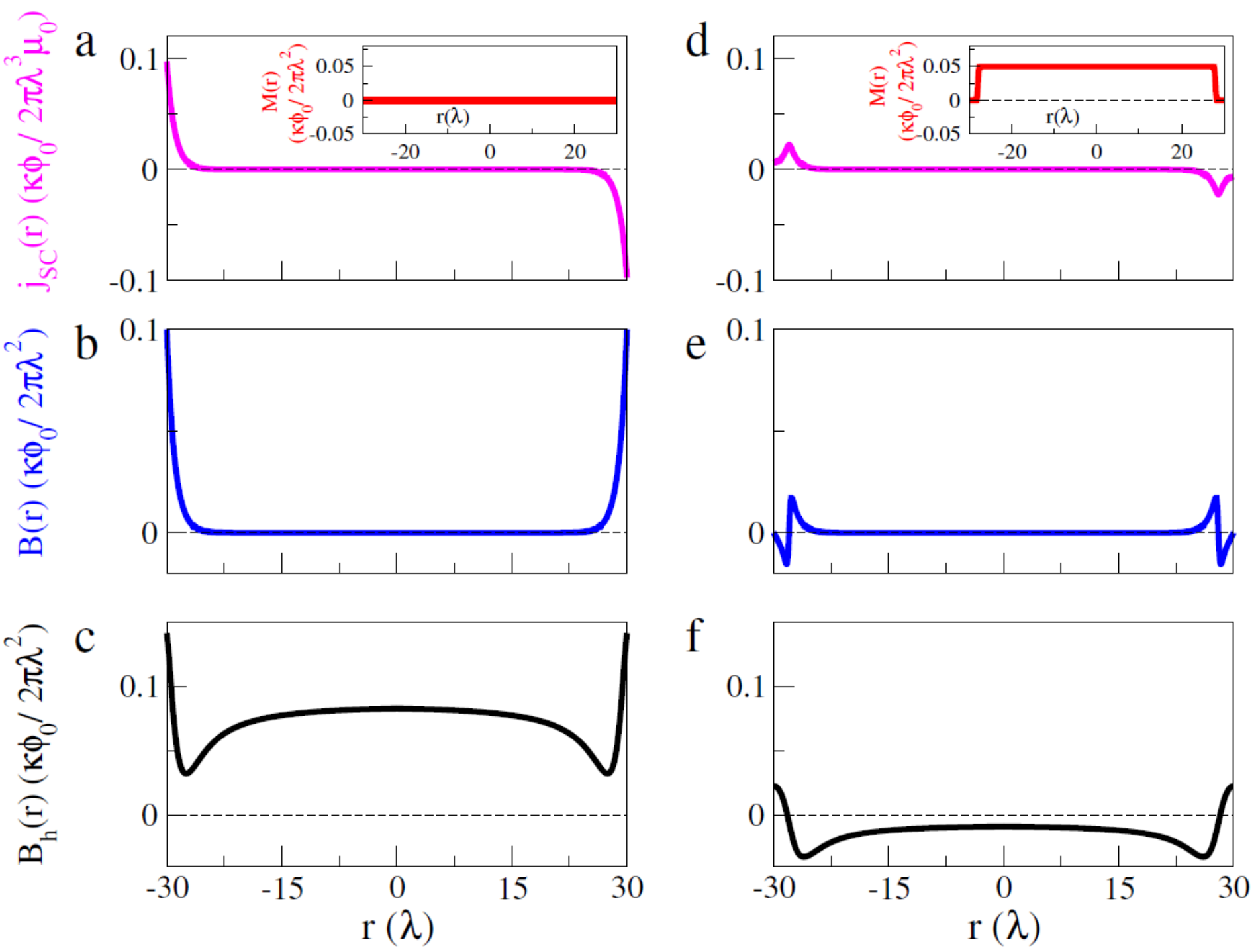

**Supplementary Fig. S7 | The current and magnetic field distribution in the transverse plane of a superconducting cylinder. a–c** Results for an externally applied field $B_{ext} = 0.1\ \kappa\phi_0/2\pi\lambda^2$ in the absence of internal magnetization. **d–f** Results without any externally applied field in the presence of an internal magnetization profile. **a, d** The in-plane current distributions, with corresponding magnetization profiles shown in the inset. **b, e** In-plane magnetic field distributions (also shown in Fig. 3 of the main text for $r_b = 10\lambda$). **c, f** The in-plane magnetic field distributions evaluated at a height $h = \lambda$ along vertical direction above the superconducting plane.

## Subsection S7.3 - Corrections due to finite sample thickness

In this subsection, we show that the inability of a superconducting cylinder with static internal magnetization to reproduce the measured remanent signal is not a consequence of the finite thickness of the sample. We consider a superconducting disc of thickness $d$, with in-plane vector potential $\vec{A}(\vec{r}_{\parallel})$. An in-plane magnetization profile in the sample $\vec{j}_M(\vec{r}_{\parallel})$ induces a transverse current density given by

$$\vec{j}_M\left(\vec{r}_{\parallel}\right) = \left(\vec{\nabla} \times \vec{M}\left(\vec{r}_{\parallel}\right)\right)_{\vec{r}_{\parallel}}. \tag{16}$$

For a small thickness $d \ll \lambda$, the current profile is uniform along the longitudinal direction. Consequently, the superconducting current density in the disc can be expressed in terms of the thin sheet current as

$$\vec{j}_S(\vec{r}) = \frac{\vec{A}(\vec{r})}{\lambda^2} \approx \frac{d}{\lambda^2}\vec{A}(\vec{r}_{||})\delta_z .$$

(17)

Similarly, the current density induced by the magnetization assumes the form

$$\vec{j}_M(\vec{r}) = \alpha d\vec{j}_M(\vec{r}_{||})\delta_z ,$$

(18)

Where $\alpha = O(1)$ encodes whatever geometric factors differentiate between the $z$ axis profile of the two current densities. Using the second GL equation, we can write

$$\nabla^2\vec{A}(\vec{r}) = \vec{j}_M(\vec{r}) - \vec{j}_S(\vec{r}) .$$

(19)

Taking the Fourier transform of Eq. (19) yields

$$- q^2\vec{A}(\vec{q}_{||}, q_z) = \frac{\alpha d}{2\pi}\vec{j}_M(\vec{q}_{||}) - \frac{d}{2\pi\lambda^2}\vec{A}(\vec{q}_{||}) .$$

(20)

Here, we have used the London relation $\vec{j}_S = \lambda^{-2}\vec{A}$, and $\chi(\vec{q}_t) = \int(dq_z/2\pi)\chi(\vec{q})$. Now we integrate Eq. (20) over $q_z$ which leads to

$$\int \vec{A}(\vec{q}_{||}, q_z)dq_z = \left(\frac{\alpha d}{2\pi}\vec{j}_M(\vec{q}_{||}) - \frac{d}{2\pi\lambda^2}\int \vec{A}(\vec{q}_{||}, q_z^{'})dq_z^{'}\right)\int \frac{-1}{q_{||}^2+q_z^2} dq_z .$$

(21)

Solving for the superconducting in-plane current profile $\vec{A}(\vec{q}_t) = \int\vec{A}(\vec{q}_t, q_z)dq_z$, we obtain

$$\vec{A}(\vec{q}_{||}) = \frac{\alpha d}{2|q_{||}|} \frac{\vec{j}_M(\vec{q}_{||})}{1+\frac{d}{2|q_{||}|\lambda^2}} .$$

(22)

The penetration depth in our system is $\lambda\approx500$ nm, and the disc thickness is $d\approx100$ nm. The corresponding ``Pearl length'' is therefore $\lambda_p = \lambda^2/d\approx2.5$ μm. When the transverse length scale (the radius of the disk in cylindrical coordinates), $R_{||}\sim2\pi/q_{||}$, is much larger than ``Pearl'' length, i.e., $R_{||}/4\pi\gg\lambda_p$, Eq. (22) reduces to

$$\vec{A}(\vec{q}_{||}) = \alpha\lambda^2\vec{j}_M(\vec{q}_{||}) .$$

(23)

In the calculations discussed in Supplementary Section S7.1, we consider $\lambda\sim1$, which implies that $\alpha\lambda^2\approx O(1)$. As a result, the contribution from the magnetization profile nearly cancels the superconducting current in the bulk, in contradiction with the finite remanent flux observed in the interior of the sample.